\documentclass[twocolumn, prb, showpacs, superscriptaddress, amsmath, amsfonts, amssymb]{revtex4}
\usepackage{graphicx}

\DeclareMathOperator{\trace}{Tr}
\newcommand{\nn}[0]{\ensuremath \nonumber \\}
\newcommand{\blr}[1]{\ensuremath \!\left( #1 \right)}
\newcommand{\bclr}[1]{\ensuremath \! \left[ #1 \right]}
\newcommand{\bslr}[1]{\ensuremath \! \left\lbrace #1 \right\rbrace}
\newcommand{\abs}[1]{\ensuremath \left| #1 \right|}
\newcommand{\ahalf}[0]{\ensuremath \frac{1}{2}}
\newcommand{\ihalf}[0]{\ensuremath \frac{i}{2}}
\renewcommand{\vec}[1]{\ensuremath \boldsymbol{ #1 } }

\newcommand{\set}[1]{\ensuremath \Bigl\lbrace #1 \Bigr\rbrace  }
\newcommand{\co}[1]{\ensuremath \cos \blr{ #1 } }
\newcommand{\si}[1]{\ensuremath \sin \blr{ #1 } }
\newcommand{\ta}[1]{\ensuremath \tan \blr{ #1 } }
\newcommand{\cosq}[1]{\ensuremath \cos^2 \blr{ #1 } }
\newcommand{\sisq}[1]{\ensuremath \sin^2 \blr{ #1 } }
\newcommand{\bra}[1]{\ensuremath \left\langle #1 \right| }
\newcommand{\ket}[1]{\ensuremath \left| #1 \right\rangle }
\newcommand{\anc}[0]{\ensuremath \hat{ c } }
\newcommand{\crc}[0]{\ensuremath \hat{ c }^{ \dagger } }
\newcommand{\anpsi}[0]{\ensuremath \hat{ \psi } }
\newcommand{\crpsi}[0]{\ensuremath \hat{ \psi }^{ \dagger } }
\newcommand{\Ha}[0]{\ensuremath \hat{ \mathcal{H} } }

\begin{document}

\title{ Non-collinear spin-spiral phase for the uniform electron gas within\\ Reduced-Density-Matrix-Functional Theory}
\author{ F.~G.~Eich }
\email[Electronic address: ]{ eich@physik.fu-berlin.de }
\affiliation{ Fritz-Haber-Institut der Max-Planck-Gesellschaft, Berlin }
\affiliation{ Institut f\"ur Theoretische Physik, Freie Universit\"at Berlin, Arnimallee 14, D-14195 Berlin }
\affiliation{ European Theoretical Spectroscopy Facility (ETSF) }
\author{ S.~Kurth }
\affiliation{ Nano-Bio Spectroscopy Group, Dpto. de F\'{i}sica de Materiales, 
Universidad del Pa\'{i}s Vasco UPV/EHU, Centro Mixto CSIC-UPV/EHU, 
Av. Tolosa 72, E-20018 San Sebasti\'{a}n, Spain }
\affiliation{ IKERBASQUE, Basque Foundation for Science, E-48011 Bilbao, Spain}
\affiliation{ European Theoretical Spectroscopy Facility (ETSF) }
\author{ C.~R.~Proetto }
\altaffiliation[Permanent address: ]{ Centro At\'omico Bariloche and Instituto Balseiro, 8400 S. C. de Bariloche, R\'io Negro, Argentina }
\affiliation{ Institut f\"ur Theoretische Physik, Freie Universit\"at Berlin, Arnimallee 14, D-14195 Berlin }
\affiliation{ European Theoretical Spectroscopy Facility (ETSF) }
\author{ S.~Sharma }
\affiliation{ Fritz-Haber-Institut der Max-Planck-Gesellschaft, Berlin }
\affiliation{ Institut f\"ur Theoretische Physik, Freie Universit\"at Berlin, Arnimallee 14, D-14195 Berlin }
\affiliation{ European Theoretical Spectroscopy Facility (ETSF) }
\author{ E.~K.~U.~Gross }
\affiliation{ Institut f\"ur Theoretische Physik, Freie Universit\"at Berlin, Arnimallee 14, D-14195 Berlin }
\affiliation{ European Theoretical Spectroscopy Facility (ETSF) }

\date{\today}

\begin{abstract}
The non-collinear spin-spiral density wave of the uniform electron gas 
is studied in the framework of Reduced-Density-Matrix-Functional Theory.
For the Hartree-Fock approximation, which can be obtained as a 
limiting case of Reduced-Density-Matrix-Functional Theory, Overhauser showed a long time ago that the 
paramagnetic state of the electron gas is unstable with respect to the 
formation of charge or spin density waves. Here we not only present a detailed 
numerical investigation of the spin-spiral density wave in the Hartree-Fock approximation 
but also investigate the effects of correlations on the spin-spiral density wave instability by 
means of a recently proposed density-matrix functional. 
\end{abstract}


\maketitle

\section{Introduction}
For many decades, the uniform electron gas
has served as the model for the description of many-particle systems 
\cite{GiulianiVignale:05}.  
However, the determination of its ground state, without any symmetry 
assumptions, still remains a challenge. Specific symmetries for the fully correlated 
uniform electron gas have been investigated using Monte Carlo methods 
\cite{CeperleyAlder:80,OrtizBallone:99}. These studies focus mostly on broken 
spatial symmetry, i.e.~, Wigner crystallization, or broken global spin symmetry. 

For the electron gas with constant electron density and uniform 
spin-polarization, the ground-state energy is analytically accessible
in the Hartree-Fock approximation. Overhauser showed in his 
seminal work \cite{Overhauser:60,Overhauser:62} that within the Hartree-Fock approximation
the aforementioned homogeneous ground state exhibits an instability w.r.t.~the
formation of charge and spin density waves.
Wigner crystallization within Hartree-Fock has been 
investigated in Ref.~\onlinecite{TrailNeeds:03}. 
Only recently the combined local spatial- and spin-symmetry breaking of the 
Hartree-Fock ground state has been studied using a Monte Carlo method 
which optimizes the ground-state energy in the space of single Slater-determinants 
\cite{ZhangCeperley:08}. However, this study still remains in the regime of 
collinear spin polarization. 

In the present work we investigate the case of local spin symmetry breaking, 
specifically a \emph{non-collinear} spin-spiral symmetry. We employ 
Reduced-Density-Matrix-Functional Theory both in the limiting case 
of the Hartree-Fock approximation as well as for the correlated electron gas using 
the recently proposed density-matrix-power functional
\cite{SharmaGross:08,LathiotakisGross:09}.

\section{Theoretical framework}
\subsection{Reduced-Density-Matrix-Functional Theory}
The basic variable in Reduced-Density-Matrix-Functional Theory (RDMFT) is the one-body-reduced density matrix (1-RDM) 
defined by
\begin{equation} \label{def_1rdm}
\gamma_{\sigma \sigma^{\prime}}\blr{\vec{r};\vec{r}^{\prime}} \equiv
\trace_N \bslr{ \hat{D} \crpsi_{\sigma^{\prime}}\blr{\vec{r}^{\prime}} 
\anpsi_{\sigma}\blr{\vec{r}} }
\end{equation}
where $ \hat{D} $ is the zero-temperature statistical operator of an ensemble of $N$-electron states
\begin{equation}
\hat{D} \equiv \sum_i \omega_i^2 \ket{\Psi^{N}_i} \bra{\Psi^{N}_i} \;\; 
\mbox{with} \;\; \sum_i \omega_i^2 = 1,
\end{equation}
where ${\crpsi_{\sigma}\blr{\vec{r}}}$ and ${\anpsi_{\sigma}\blr{\vec{r}}}$ 
are fermionic creation and annihilation operators, respectively.
The 1-RDM is a Hermitian operator in the single-particle Hilbert space and 
can be represented by its spectral decomposition 
\begin{equation} \label{1rdm}
\gamma\blr{\vec{r};\vec{r}^{\prime}} = 
\sum_i n_i \Phi_{i}\blr{\vec{r}} \Phi^{\dagger}_{i}\blr{\vec{r}^{\prime}},
\end{equation}
where the eigenvalues $ n_i $ are called occupation numbers (ON) and the 
corresponding single-particle Pauli-spinor eigenstates 
$ \Phi_i(\vec{r}) = \left( \varphi_{i \uparrow}\blr{\vec{r}}, 
\varphi_{i \downarrow}\blr{\vec{r}} \right)^T$ 
are referred to as natural orbitals (NO). It was shown by Gilbert 
\cite{Gilbert:75} that the $N$-particle ground state is a unique functional 
of the ground state 1-RDM, i.e.~, $\ket{\Psi^{N}_0}= \ket{\Psi^{N}_0 
\bclr{\gamma^{\mathrm{gs}}}}$ . Therefore the ground state energy for a 
system of $N$ interacting electrons moving in an arbitrary but fixed 
(possibly non-local) external potential $\hat{V}$ is also a functional of the 1-RDM:
\begin{equation} \label{E_gs_def}
E_{\mathrm{V}} \bclr{\gamma^{\mathrm{gs}}} = \bra{\Psi^{N}_0 
\bclr{\gamma^{\mathrm{gs}}}} \Ha_V \ket{\Psi^{N}_0 
\bclr{\gamma^{\mathrm{gs}}}}  \;,
\end{equation}
where ${ \Ha_V = \hat{T} + \hat{V} + \hat{W} + E_{\mathrm{ion}} }$ is a generic 
interacting many-body Hamiltonian with kinetic energy $ \hat{T} $, external 
potential $ \hat{V} $, electron-electron interaction $ \hat{W} $, and 
a constant energy contribution $ E_{\mathrm{ion}} $ from the degrees of freedom
that are not treated quantum mechanically. 

The ground-state-energy functional can be decomposed into the following 
components
\begin{equation} \label{E_gs}
E_{\mathrm{V}} \bclr{\gamma} = 
T\bclr{\gamma} + V\bclr{\gamma} + W\bclr{\gamma} + E_{\mathrm{ion}},
\end{equation}
with the kinetic energy (atomic units are used throughout the paper, and the 
superscript ``gs'' is omitted for brevity)
\begin{equation} \label{T}
T\bclr{\gamma} = \sum_{\sigma} \int\!\!\mathrm{d}^3r 
\lim_{\vec{r}^{\prime}\to\vec{r}} \ahalf \nabla^{\prime} \nabla
\gamma_{\sigma \sigma}\blr{\vec{r};\vec{r}^{\prime}},
\end{equation}
and the energy contribution due to the external potential
\begin{equation} \label{V}
V\bclr{\gamma} = \sum_{\sigma} \int\!\!\mathrm{d}^3r 
V\blr{\vec{r}} \gamma_{\sigma \sigma}\blr{\vec{r};\vec{r}}.
\end{equation}
Here we are assuming a local, spin independent external potential.
The Hohenberg-Kohn theorem of Density-Functional Theory (DFT) proves a 
one-to-one mapping between the ground-state density
and the N-particle ground state, considering only local external potentials. However, in 
RDMFT the Gilbert theorem ensures a one-to-one correspondence
between the ground-state 1-RDM and the N-particle ground state by considering the broader 
class of non-local external potentials. This also implies the
one-to-one mapping between a local potential and the ground-state 1-RDM. Note that in contrast to usual
Kohn-Sham DFT \emph{all} single-particle contributions to the ground state energy $ E_{\mathrm{V}} $ are explicitly given
in terms of the ground state 1-RDM.
However, the interaction energy
\begin{equation} \label{W}
W\bclr{\gamma}= \sum_{\sigma_1 \sigma_2} \iint\!\!\mathrm{d}^3r_1 \mathrm{d}^3r_2 
\frac{P^{\mathrm{gs}}_{\sigma_1 \sigma_2} \bclr{\gamma} \blr{\vec{r}_1, \vec{r}_2}}{\abs{\vec{r}_1-\vec{r}_2}},
\end{equation}
is only known explicitly in terms of the ground-state pair density
\begin{align} \label{def_pair_denisty}
\lefteqn{P^{\mathrm{gs}}_{\sigma_1 \sigma_2} \bclr{\gamma} 
\blr{\vec{r}_1, \vec{r}_2} \equiv } \nn 
&\bra{\Psi^{N}_0 \bclr{\gamma}} 
\crpsi_{\sigma_1}\blr{\vec{r}_1} \crpsi_{\sigma_2}
\blr{\vec{r}_2} \anpsi_{\sigma_2}\blr{\vec{r}_2}
\anpsi_{\sigma_1}\blr{\vec{r}_1} \ket{\Psi^{N}_0 \bclr{\gamma}} \; .
\end{align}

The basic idea of RDMFT is to extend the domain of the ground-state-energy functional in 
Eq.~\eqref{E_gs_def} to all ensemble-$N$-representable 1-RDMs [as defined in Eq.~\eqref{def_1rdm}] and then 
employ the variational principle in order to find the ground-state 1-RDM as 
well as the ground-state energy corresponding to a fixed external potential $V$. 
The necessary and sufficient conditions for a 1-RDM to be 
ensemble-$N$-representable are \cite{Coleman:63}:
\begin{subequations}
\label{nrep}
\begin{align} 
& \sum_i n_i = N \;\; \mbox{and} \;\; 0 \leq n_i \leq 1 , \label{nrep_on} \\
& \sum_{\sigma} \int\!\!\mathrm{d}^3r \varphi^{\star}_{i \sigma}\blr{\vec{r}}
\varphi_{j \sigma}\blr{\vec{r}} = \delta_{ij}. \label{nrep_no}
\end{align}
\end{subequations}
In order to apply RDMFT in practice we need to approximate the functional 
dependence of the pair density on the 1-RDM. Since we want to study the 
spin-spiral density wave (SSDW) instability in the uncorrelated (Hartree-Fock, 
HF) and the correlated regime, we focus on the so-called 
density-matrix-power functional introduced in Ref.~\onlinecite{SharmaGross:08}:
\begin{align} \label{pf_def}
P^{\alpha}_{\sigma_1 \sigma_2} \bclr{\gamma} \blr{\vec{r}_1, \vec{r}_2} \equiv & 
\ahalf \gamma_{\sigma_1 \sigma_1}\blr{\vec{r}_1;\vec{r}_1} \gamma_{\sigma_2 \sigma_2}\blr{\vec{r}_2;\vec{r}_2} \nn
& - \ahalf \gamma^{\alpha}_{\sigma_1 \sigma_2}\blr{\vec{r}_1;\vec{r}_2} 
\gamma^{\alpha}_{\sigma_2 \sigma_1}\blr{\vec{r}_2;\vec{r}_1} \; ,
\end{align}
for ${ 0.5 \leq \alpha \leq 1 }$. Here the power of the 1-RDM has to be read 
in the operator sense, i.e.
\begin{equation} 
\gamma^{\alpha} \blr{\vec{r};\vec{r}^{\prime}} = 
\sum_i n_i^{\alpha}  \Phi_{i}\blr{\vec{r}} \Phi^{\dagger}_{i}\blr{\vec{r}^{\prime}}.
\end{equation}
As limiting cases it contains both the uncorrelated HF approximation (for ${\alpha=1}$) as 
well as the correlated M\"uller or Buijse-Baerends
functional (for ${\alpha=0.5}$)\cite{Mueller:84,BuijseBaerends:02}.
Also, it was recently shown\cite{LathiotakisGross:09} 
that the power functional yields good correlation energies for the unpolarized 
uniform electron gas. 

\subsection{The Overhauser Instability of the uniform electron gas}
\begin{figure}
\includegraphics[width=\columnwidth]{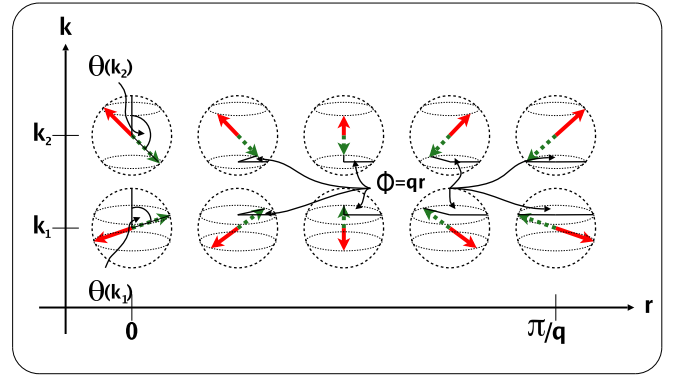}
\caption{\label{sdw_rotation} 
(Color online) The effects of the spin rotation ${\mathcal{U}\blr{\vec{r};\vec{k}}}$
on pure spin-up (dashed arrow) or pure spin-down (solid arrow) natural orbitals (plane waves) for
two momenta $\vec{k}_{1/2}$. The angle $\theta_{\vec{k}}$ specifies the cone on which the spin is rotating.
The position on the cone is given by the angle $\phi\blr{\vec{r}}=\vec{q}\!\cdot\!\vec{r}$, which is the same for all
natural orbitals.} 
\end{figure}

The system under investigation is the uniform electron gas (UEG) in three 
dimensions, i.e.~, a gas of interacting electrons subject to an external 
potential induced by a uniformly distributed positive background charge. 
Overhauser has proved that the true HF 
ground state does not correspond to a homogeneous electron density (although 
there are solutions to the HF equations where the symmetry is not broken), 
since the HF energy can be lowered by forming a charge density wave (CDW) or spin density wave (SDW)
\cite{Overhauser:62}. As an explicit example he assumed, in addition to the regular HF 
potential $ V_{\sigma \vec{k}} $, a potential $ g_{\vec{k}} $ in the 
HF Hamiltonian that couples plane waves of opposite spin whose momenta 
differ by $ \vec{q} $:
\begin{align}
\Ha^{\mathrm{HF}} = & \sum_{\vec{k} \sigma} \bslr{\frac{k^2}{2}-V_{\sigma \vec{k}}} \crc_{\vec{k} \sigma} \anc_{\vec{k} \sigma} \nn
& - \sum_{\vec{k}} g_{\vec{k}}\bslr{ \crc_{\vec{k}+\frac{\vec{q}}{2} \uparrow} \anc_{\vec{k}-\frac{\vec{q}}{2} \downarrow}
+ \crc_{\vec{k}-\frac{\vec{q}}{2} \downarrow} \anc_{\vec{k}+\frac{\vec{q}}{2} \uparrow}}. \label{ssw_hf}
\end{align}
Overhauser demonstrated that with the ansatz 
\begin{subequations}
\label{no_def}
\begin{align} 
\Phi_{1 \vec{k}}\blr{\vec{r}} = & 
\begin{pmatrix} 
\Big. \co{\ahalf\theta_{\vec{k}}}e^{-\ihalf \vec{q}\cdot\vec{r}} \\
\Big. \si{\ahalf\theta_{\vec{k}}}e^{\ihalf \vec{q}\cdot\vec{r}} \end{pmatrix} 
\frac{e^{i \vec{k}\cdot\vec{r}}}{\sqrt{\Omega}} , \label{no_b1}\\
\Phi_{2 \vec{k}}\blr{\vec{r}} = & 
\begin{pmatrix} 
\Big. -\si{\ahalf\theta_{\vec{k}}}e^{-\ihalf \vec{q}\cdot\vec{r}} \\
\Big. \co{\ahalf\theta_{\vec{k}}}e^{\ihalf \vec{q}\cdot\vec{r}} \end{pmatrix}
\frac{e^{i \vec{k}\cdot\vec{r}}}{\sqrt{\Omega}}, \label{no_b2}
\end{align}
\end{subequations}
the HF self-consistent equations are transformed into a set of equations
relating the orbital angles $\theta_{\vec{k}} $, the potential 
$ g_{\vec{k}} $ and the regular HF potential $ V_{\sigma \vec{k}} $. 
Note that the generic single-particle index $i$ here has been replaced by the 
joint index ${i\to\bslr{\blr{b=1,2},\vec{k}}}$.
After taking the thermodynamic limit 
( i.e.~, the volume $\Omega$ and the number of particles $N$ are taken to be 
infinity such that ${\frac{N}{\Omega}}$ remains constant), these equations 
read (cf.~Ref.~\onlinecite{GiulianiVignaleSDW:05})
\begin{subequations}
\label{ss_hf_sc}
\begin{align}
V_{\uparrow \vec{k}-\frac{\vec{q}}{2}} = & 
\int\!\! \frac{\mathrm{d}^3k^{\prime}}{\blr{2 \pi}^3} \frac{4 \pi}{\abs{\vec{k}-\vec{k}^{\prime}}^2} \nn
& \times \bslr{n_{1 \vec{k}^{\prime}} \cosq{\frac{\theta_{\vec{k}^{\prime}}}{2}}
+n_{2 \vec{k}^{\prime}} \sisq{\frac{\theta_{\vec{k}^{\prime}}}{2}}}, \hphantom{xx} \label{v_up_sc} \\
V_{\downarrow \vec{k}+\frac{\vec{q}}{2}} = & 
\int\!\! \frac{\mathrm{d}^3k^{\prime}}{\blr{2 \pi}^3} \frac{4 \pi}{\abs{\vec{k}-\vec{k}^{\prime}}^2} \nn
& \times \bslr{n_{1 \vec{k}^{\prime}} \sisq{\frac{\theta_{\vec{k}^{\prime}}}{2}}
+n_{2 \vec{k}^{\prime}} \cosq{\frac{\theta_{\vec{k}^{\prime}}}{2}}}, \hphantom{xx} \label{v_down_sc} \\
2 g_{\vec{k}} = & 
\int\!\! \frac{\mathrm{d}^3k^{\prime}}{\blr{2 \pi}^3} \frac{4 \pi}{\abs{\vec{k}-\vec{k}^{\prime}}^2} 
\bslr{n_{1 \vec{k}^{\prime}} - n_{2 \vec{k}^{\prime}} } \si{\theta_{\vec{k}^{\prime}}} . \label{g_sc}
\end{align}
\end{subequations}
The r.h.s.~of Eqs.~\eqref{ss_hf_sc} implicitly depends on $\vec{q}$ via the ${n_{b \vec{k}}}$ and the ${\theta_{\vec{k}}}$.
The $ n_{b \vec{k}} $ are the occupation numbers (either 0 or 1) of the 
orbitals $ \Phi_{b \vec{k}} $ which comprise the HF ground-state 
Slater-determinant and specify the Fermi surface (the boundaries of the integration) in Eqs.~(\ref{v_up_sc}) - (\ref{g_sc}). 
The orbital angles $ \theta_{\vec{k}} $ on the other hand  are given by
\begin{subequations}
\label{theta_pot}
\begin{align}
\ta{\theta_{\vec{k}}} = & \frac{2 g_{\vec{k}}}{\epsilon_{\uparrow \vec{k}-\frac{\vec{q}}{2}} - \epsilon_{\downarrow \vec{k}+\frac{\vec{q}}{2}}}, \label{sc_theta} \\
\epsilon_{\uparrow \vec{k}-\frac{\vec{q}}{2}} = & \frac{\blr{\vec{k}-\ahalf\vec{q}}^2}{2} - V_{\uparrow \vec{k}-\frac{\vec{q}}{2}}, \label{e_up_def} \\
\epsilon_{\downarrow \vec{k}+\frac{\vec{q}}{2}} = & \frac{\blr{\vec{k}+\ahalf\vec{q}}^2}{2} - V_{\downarrow \vec{k}+\frac{\vec{q}}{2}}. \label{e_down_def}
\end{align}
\end{subequations}
Note that the origin in momentum space is shifted by $ \vec{q}/2 $ compared to 
the definitions in Ref.~\onlinecite{GiulianiVignaleSDW:05}.
The energy contribution due to the pairing potential $ g_{\vec{k}} $ favors 
a hybridization of spin-up and spin-down plane waves differing by $ \vec{q} $ 
in their momenta. The orbital angles ${\theta_{\vec{k}}}$ introduced in Overhauser's ansatz Eq.~\eqref{no_def}
describe this hybridization. Another way of looking at the orbital angles ${\theta_{\vec{k}}}$ is to consider them, together
with the angles ${\phi\blr{\vec{r}}=\vec{q}\!\cdot\!\vec{r}}$,
as angles defining a rotation in spin space represented by
\begin{align}
& \mathcal{U}\blr{\vec{r};\vec{k}} \equiv e^{-\imath \phi\blr{\vec{r}} \sigma^z} e^{-\imath \theta_{\vec{k}} \sigma^y} \nn
& = \begin{pmatrix} \Big. \co{\ahalf\theta_{\vec{k}}}e^{-\ihalf \vec{q} \cdot \vec{r}} &
-\si{\ahalf\theta_{\vec{k}}}e^{-\ihalf \vec{q} \cdot \vec{r}} \\
\Big. \si{\ahalf\theta_{\vec{k}}}e^{\ihalf \vec{q} \cdot \vec{r}} &
\co{\ahalf\theta_{\vec{k}}}e^{\ihalf \vec{q} \cdot \vec{r}} \end{pmatrix} , \label{Udef}
\end{align}
where $ \sigma^{y/z} $ are Pauli matrices. The orbitals of Eq.~\eqref{no_b1}[\eqref{no_b2}] can then 
be thought of as being constructed by transforming 
pure spin-up [spin-down] plane waves in spin space according to the rotation Eq.~\eqref{Udef}.
First the plane wave is rotated 
around the $y$-axis by an angle $ \theta_{\vec{k}} $, i.e.~, an angle depending 
on its momentum. Then it is rotated around the $z$-axis by 
an angle ${ \phi\blr{\vec{r}}=\vec{q}\!\cdot\!\vec{r} }$ which is the same 
for all plane waves, independent of the wave vector, but depends on the 
spatial position (see Fig.~\ref{sdw_rotation}). With this consideration it is 
clear that the angle $ \theta_{\vec{k}} $ has to be restricted to the interval 
${ \bclr{0,\pi} }$ in order to assign a unique azimuthal rotation 
angle.

In previous studies 
within RMDFT \cite{LathiotakisGross:07,LathiotakisGross:09} it was assumed that
the 1-RDM exhibits the symmetries present in the Hamiltonian, i.e.~, the NOs are 
pure spin-up(down) plane waves, while here
we use orbitals of the form of Eq.~(\ref{no_def}) as NOs
for our RDMFT treatment of the UEG. 
The spin-spiral wave vector $ \vec{q} $ and the angle $ \theta_{\vec{k}} $ will be treated as variational 
parameters for the NOs. 
It can easily be verified that the NOs of Eq.~(\ref{no_def}) form a complete 
and orthonormal set and that the corresponding electron density ${\rho \equiv 3/\blr{4 \pi r_s^3}}$,
given in terms of the Wigner-Seitz radius $r_s$, is still spatially constant.
The magnetization of the UEG is defined by
\begin{align} 
\vec{m}\blr{\vec{r}} \equiv & - \ahalf \sum_{\sigma \sigma^{\prime}}\bra{\Psi} \crpsi_{\sigma^{\prime}}\blr{\vec{r}}
\vec{\sigma}_{\sigma \sigma^{\prime}} \anpsi_{\sigma}\blr{\vec{r}} \ket{\Psi} \label{mag_def} \\ 
= & - \begin{pmatrix} \mathcal{R} \gamma_{\uparrow \downarrow}\blr{\vec{r};\vec{r}} \\ 
\mathcal{I} \gamma_{\uparrow \downarrow}\blr{\vec{r};\vec{r}} \\ 
\ahalf \bslr{\gamma_{\uparrow \uparrow}\blr{\vec{r};\vec{r}} - \gamma_{\downarrow \downarrow}\blr{\vec{r};\vec{r}}}\end{pmatrix} , \nonumber
\end{align}
and varies in space as
\begin{subequations}
\label{magnetization}
\begin{align}
\vec{m}\blr{\vec{r}} = &  
- \begin{pmatrix} A \co{\vec{q}\!\cdot\!\vec{r}} \\ A \si{\vec{q}\!\cdot\!\vec{r}} \\ B \end{pmatrix}, \label{m_def} \\
A \equiv & \ahalf \int\!\! \frac{\mathrm{d}^3k}{\blr{2 \pi}^3} 
\bslr{n_{1 \vec{k}}-n_{2 \vec{k}}} \si{\theta_{\vec{k}}} , \label{amp_xy} \\
B \equiv & \ahalf \int\!\! \frac{\mathrm{d}^3k}{\blr{2 \pi}^3} 
\bslr{n_{1 \vec{k}}-n_{2 \vec{k}}} \co{\theta_{\vec{k}}} , \label{amp_z} 
\end{align}
\end{subequations}
i.e.~, the $ x $- and $ y $-components of the magnetization rotate in space along the direction of $ \vec{q} $ with
a periodicity given by the wavelength ${ q = \abs{\vec{q}} }$. This geometry of the magnetization is
usually referred to as SSDW \footnote{In Ref.~\onlinecite{ZhangCeperley:08}
the x- and y-components of the magnetization are locally zero and the z-component varies in space, such that 
its global value is also zero (collinear configuration).}.

\section{Numerical Implementation} \label{sec_num}
Having chosen a functional and having made an ansatz for the NOs, we
minimize the functional for the ground-state energy. The functional depends on 
$ n_{b \vec{k}} $ , $ \theta_{\vec{k}} $ and the spin-spiral wave vector 
$\vec{q}$. The contribution $ E_{\mathrm{ion}} $ coming from the uniform 
positive background charge cancels exactly the classical contribution of the 
interaction energy, since the density is constant. Accordingly the energy per electron reads
\begin{equation} 
e_{\alpha}\bclr{n_b, \theta}\blr{\vec{q}} = t\bclr{n_b, \theta}\blr{\vec{q}}
 - w_{\alpha 1}\bclr{n_b, \theta} - w_{\alpha 2}\bclr{n_b, \theta} \; , 
\label{energy_functional}
\end{equation}
with the kinetic energy per electron 
\begin{align}
t\bclr{n_b, \theta}\blr{\vec{q}} = & 
\frac{1}{2 \rho} \int\!\! \frac{\mathrm{d}^3k}{\blr{2 \pi}^3} \Big\lbrace\! \blr{n_{1\vec{k}}+n_{2\vec{k}}}k^2 \nn
& - \vec{q}\!\cdot\!\vec{k} \blr{n_{1\vec{k}}-n_{2\vec{k}}}
\co{\theta_{\vec{k}}} \!\Big\rbrace + \frac{q^2}{8} \; , \label{e_kin}
\end{align}
the energy contribution from exchange-like terms of orbitals with the same $b$ (intra-band exchange)
\begin{align}
w_{\alpha 1}\bclr{n_b, \theta} = & \frac{1}{2 \rho} \iint\!\! \frac{\mathrm{d}^3k_1 \mathrm{d}^3k_2}{\blr{2 \pi}^6}
\frac{4 \pi}{\blr{\vec{k}_1-\vec{k}_2}^2} \nn 
& \times \bslr{\blr{n_{1\vec{k}_1} n_{1\vec{k}_2}}^{\alpha} + \blr{n_{2\vec{k}_1} n_{2\vec{k}_2}}^{\alpha}} \nn
& \times \cosq{\frac{\theta_{\vec{k}_1}-\theta_{\vec{k}_2}}{2}} \; , 
\label{e_intra_b}
\end{align}
and the energy contribution from exchange-like terms of orbitals with opposite $b$ (inter-band exchange)
\begin{align}
w_{\alpha 2}\bclr{n_b, \theta} = & \frac{1}{2 \rho} \iint\!\! \frac{\mathrm{d}^3k_1 \mathrm{d}^3k_2}{\blr{2 \pi}^6}
\frac{4 \pi}{\blr{\vec{k}_1-\vec{k}_2}^2} \nn
& \times \bslr{\blr{n_{1\vec{k}_1} n_{2\vec{k}_2}}^{\alpha} + \blr{n_{2\vec{k}_1} n_{1\vec{k}_2}}^{\alpha}} \nn
& \times \sisq{\frac{\theta_{\vec{k}_1}-\theta_{\vec{k}_2}}{2}} \; . 
\label{e_inter_b}
\end{align}
We assume that the symmetry is only broken along the direction of $\vec{q}$
which is chosen to be parallel to the $ z $-axis.
Accordingly we can use cylindrical coordinates in momentum space, i.e.~, ${n_{b \vec{k}}=n_{b k_{\rho} k_z} }$
and ${ \theta_{\vec{k}}=\theta_{k_{\rho} k_z} }$. We also use the following additional symmetry assumptions 
\begin{subequations}
\label{sym_asumptions}
\begin{align}
n_{b k_{\rho} -k_z} = & n_{b k_{\rho} k_z} \;\; , \;\; n_{1 \vec{k}} \geq n_{2 \vec{k}}, \label{sym_n} \\
\theta_{k_{\rho} \pm \abs{k_z}} = & \frac{\pi}{2}\bslr{1 \mp a_{k_{\rho} \abs{k_z}}}, \label{sym_theta}
\end{align}
\end{subequations}
with ${ 0 \leq a_{\vec{k}} \leq 1 }$. In this way we guarantee that the energy 
gain in the part of the energy which explicitly depends on $\vec{q}$ is maximized.
The $z$-component of the magnetization vanishes under 
these symmetry assumptions (planar spiral). 

The configurations
\begin{subequations}
\label{pm_fm}
\begin{align}
n^{\mathrm{PM}}_{1\vec{k}} = & \Theta\blr{\abs{\vec{k}-\vec{e}_zk_f}-k_f}+\Theta\blr{\abs{\vec{k}+\vec{e}_zk_f}-k_f}, \nn
n^{\mathrm{PM}}_{2\vec{k}} = & 0 \; , \; a^{\mathrm{PM}}_{\vec{k}}=1 \; , \; \vec{q}=2k_f \vec{e}_z \label{pm_conf} \\
& \mathrm{and} \nn
n^{\mathrm{FM}}_{1\vec{k}} = & \Theta\blr{\abs{\vec{k}-2^{1/3}k_f}}, \nn 
n^{\mathrm{FM}}_{2\vec{k}} = & 0 \; , \; a^{\mathrm{FM}}_{\vec{k}}=0 \; , \; \vec{q}=0, \label{fm_conf} \\
k_f\equiv & \blr{\frac{9 \pi}{4} }^{1/3}\frac{1}{r_s}, \nonumber
\end{align}
\end{subequations}
which are compatible with Eq.~\eqref{sym_asumptions}, correspond to the non-magnetic (usually in this context
called paramagnetic [PM]) and ferromagnetic (FM) state of the UEG within 
HF, respectively.

When discretizing the integrals of Eqs.~(\ref{e_kin})-(\ref{e_inter_b}) we 
assume that the ONs $ n_{b \vec{k}} $ and the angles $ a_{\vec{k}} $ are 
constant within annular regions in $\vec{k}$-space
\begin{equation}
\Omega_i \equiv \set{\vec{k} \;\left\vert\; k^{i-}_{\rho} \leq k_{\rho} \leq k^{i+}_{\rho} \; ; \;
k^{i-}_z \leq k_{z} \leq k^{i+}_{z} \right. }.
\end{equation}
Then the discretized energy contributions are
\begin{subequations}
\label{disc_en}
\begin{align}
t\bclr{n_{b i}, \theta_i}\blr{q} = & \sum_{b i} n_{b i} \mathrm{DKI}_i + \frac{q^2}{8} \nn
& - q \sum_i \blr{n_{1i}-n_{2i}} \co{\theta_i} \mathrm{DQI}_i \; , \label{e_kin_disc} \\
w_{\alpha 1}\bclr{n_{b i}, \theta_i} = & \frac{1}{2} \sum_{b i j} \blr{n_{b i} n_{b j}}^{\alpha} 
\cosq{\frac{\theta_{i}-\theta_{j}}{2}} \mathrm{DXI}_{ij} \; , \;\;\;\;\;\;\; \label{e_intra_b_disc} \\
w_{\alpha 2}\bclr{n_{b i}, \theta_i} = & \sum_{i j} \blr{n_{1 i} n_{2 j}}^{\alpha} 
\sisq{\frac{\theta_{i}-\theta_{j}}{2}} \mathrm{DXI}_{ij} \; , \;\;\; \label{e_inter_b_disc}
\end{align}
\end{subequations}
where the integral weights are given by
\begin{subequations}
\begin{align}
& \mathrm{DKI}_i \equiv \frac{1}{8 \pi^2 \rho} \iint_{\Omega_i}\!\! \mathrm{d}k_{\rho} \mathrm{d}k_z
\blr{k_{\rho}^3+k_{\rho}k_z^2} \label{dki_def} \\
& \mathrm{DQI}_i \equiv \frac{1}{8 \pi^2 \rho} \iint_{\Omega_i} \!\! \mathrm{d}k_{\rho} \mathrm{d}k_z
\blr{ k_{\rho}k_z } \label{dqi_def} \\
& \mathrm{DXI}_{ij} \equiv \frac{1}{2 \rho} \iiint_{\Omega_i} \!\! 
\frac{\mathrm{d}k_{\rho1}\mathrm{d}k_{z1} \mathrm{d}\phi_1}{\blr{2\pi}^3}
\iiint_{\Omega_j} \!\! \frac{\mathrm{d}k_{\rho2} \mathrm{d}k_{z2} \mathrm{d}\phi_2}{\blr{2\pi}^3} \nn
& \times \frac{4 \pi k_{\rho1} k_{\rho2}}{k_{\rho1}^2+k_{\rho2}^2+\blr{k_{z1}-k_{z2}}^2-2k_{\rho1}k_{\rho2}
\co{\phi_1-\phi_2}} \; . \label{dxi_def}
\end{align}
\end{subequations}
The integrals \eqref{dki_def} and \eqref{dqi_def} are readily solved and the 
integrals \eqref{dxi_def} can ultimately be reduced to elliptic integrals,
which are numerically accessible with high accuracy. Since the momenta
are treated as continuous variables we stay in the thermodynamic limit.
Thus all energies obtained numerically are variational. 
The error introduced by the discretization is solely due
to the assumption that the ${n_{b\vec{k}}}$ and ${\theta_{\vec{k}}}$ are 
constant within the elementary volume elements ${\Omega_i}$ and can 
systematically be reduced by increasing the number of discretization points. 

After having discretized the problem, the minimization of the energy 
functional of Eq.~\eqref{energy_functional} becomes a high-dimensional 
optimization problem. We use a steepest descent algorithm for the minimization 
and ensure that the constraints, Eq.~\eqref{nrep}, are satisfied during the 
minimization process. Starting from some
initial 1-RDM and some initial discretization in momentum space the energy is 
minimized for a fixed spin-spiral wave vector $ \vec{q} $. Then the 
discretization is refined in those regions of momentum space where the 
$ n_{b i} $ and/or the $ a_i $ show the largest variations. 
The minimization on the refined momentum space mesh starts from a 
re-initialized 1-RDM in order to prevent dependencies on the result of the 
minimization on the coarser grid. Finally we compare the total energies at 
different $\vec{q}$ in order to determine the optimal spin-spiral
wave vector $\vec{q}_{opt}$ for various densities.

\section{Results}
\subsection{Hartree-Fock} \label{sec_hf}
\begin{figure}
\includegraphics[clip,width=0.95\columnwidth]{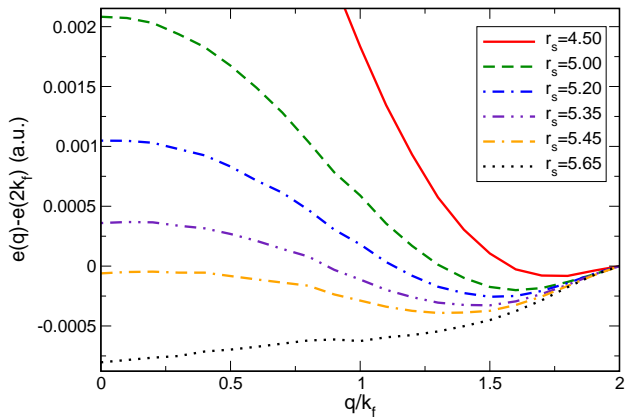}
\caption{\label{hf_e_var_rs} 
(Color online) Hartree-Fock total energy per electron of the SSDW state as function of the 
spin-spiral wave vector $ q $ at various $ r_s $. The value
$ e\blr{q=2k_f} $ is subtracted in order to emphasize the behavior of the minimum at different densities.
For increasing density (decreasing $ r_s $) the minimum shifts to higher values of $ q_{opt} $ and
the energy gained against the paramagnetic state by forming a spin spiral 
decreases.}
\end{figure}
\begin{figure}
\includegraphics[clip,width=0.95\columnwidth]{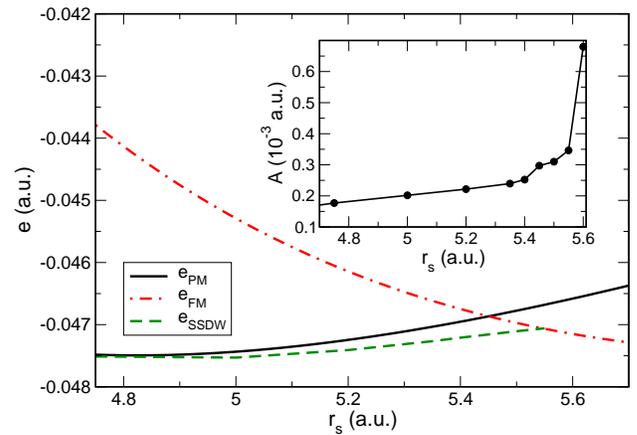}
\caption{\label{phases_amp} 
(Color online) Dependence of the energy per electron on $r_s$, within the 
Hartree-Fock approximation, for the paramagnetic, ferromagnetic and SSDW phase 
in the region of the paramagnetic-ferromagnetic crossover.
The inset shows the behavior of the amplitude $A$, defined in Eq.~\eqref{amp_xy}, at the optimal
spin-spiral wave vector.}
\end{figure}
We first use our numerical implementation to investigate Overhauser's SSDW 
state in the HF approximation, i.e.~, the density-matrix-power 
functional with ${ \alpha=1 }$. 
From the considerations in Eqs.~\eqref{pm_fm} we see that it is sufficient to 
minimize w.r.t.~a 1-RDM whose ONs are only non-zero for orbitals with ${ b = 1}$ and 
${\abs{\vec{q}} \in \bclr{0,2 k_f}}$ since both the paramagnetic and the 
ferromagnetic HF solutions are accessible under these conditions. The 
minimization at ${q=0}$ and ${q=2k_f}$ yields exactly the ONs $ n_{b i} $ and 
angle parameters $ a_i $ given in Eqs.~\eqref{fm_conf} and \eqref{pm_conf}, 
respectively. Therefore we can read the total energy per particle as function 
of the spin-spiral wave vector in the following way: $ e\blr{q=0} $ is the 
energy of the ferromagnetic state, $ e\blr{q=2 k_f} $ corresponds to the 
energy of the paramagnetic state. For intermediate values, $0 < q < 2 k_f$, 
${ e\blr{q}}$ corresponds to a SSDW configuration with ${ m_z = 0 }$ (planar spiral).
Overhauser's statement can then be expressed as ${ \left. \partial_{q} 
e\blr{q} \right|_{q=2 k_f} > 0 }$, i.e.~, the paramagnetic configuration is 
unstable w.r.t.~the formation of a SSDW. 

In Fig.~\ref{hf_e_var_rs} we show the dependence of the total energy per 
particle on the spin-spiral wave vector $\vec{q}$ for various densities.
Consistent with Overhauser's proof, the derivative of ${ e\blr{q}}$ is positive at ${q=2k_f}$.
It is clear from Fig.~\ref{hf_e_var_rs} that the optimal spin-spiral wave vector
moves away from the paramagnetic configuration ${(q=2k_f)}$ as the density decreases.
Furthermore the difference between the total energy at the 
minimum and the total energy at $q=2k_f$ increases with increasing $r_s$, i.e.~, 
the instability is more pronounced at lower densities. Below some critical 
density, however,  the ferromagnetic state ($q=0$) becomes the most stable 
solution. This is not in contradiction with Overhauser's statement since the 
spin-spiral state is still lower in energy than the paramagnetic state.
A comparison of the energy per electron in the paramagnetic, ferromagnetic and 
SSDW phase is depicted in Fig.~\ref{phases_amp}. We provide results for the non-collinear
magnetic states of the UEG in order to extend the picture given in Ref.~\onlinecite{ZhangCeperley:08}.
It seems that the gain in energy by forming a collinear SDW/CDW state as presented in Ref.~\onlinecite{ZhangCeperley:08}
is larger compared to the energy gain by forming a SSDW. This is consistent with the qualitative argument
already given by Overhauser, that the superposition of a left- and right-rotating SSDW yielding
a collinear SDW will increase the gain in energy \cite{Overhauser:62}. 

\begin{figure}
\includegraphics[clip,width=0.95\columnwidth]{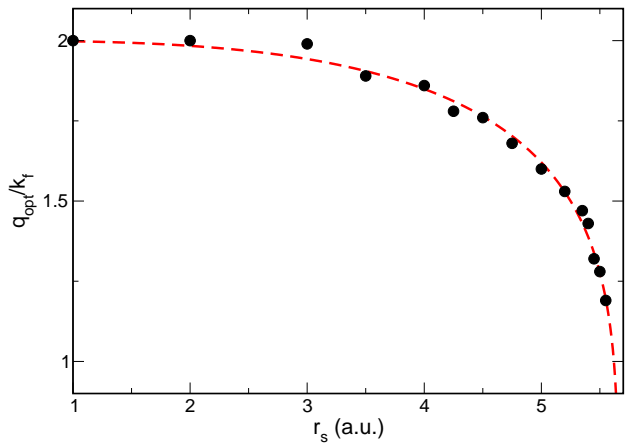}
\caption{\label{q_opt_rs} 
(Color online) Dependence of the Hartree-Fock optimal spin-spiral wave vector $q_{opt}$ on the 
density, given by $r_s$. The proposed approximation by a simple scaling law 
Eq.~\eqref{q_scaling} is shown as the dashed line.}
\end{figure}

To describe the resulting behavior of $q_{opt}\blr{r_s}$
we propose a simple, empirical scaling law for the optimal spin-spiral 
wave vector
\begin{equation} \label{q_scaling}
q_{opt}\blr{r_s}=2 k_f\blr{1-\blr{\frac{r_s}{r_0}}^3}^{\beta},
\end{equation}
where ${r_0 \approx 5.7}$ and ${\beta \approx 0.2}$.
The proposed scaling behavior of $q_{opt}$ reproduces the numerical data very accurately as can be seen
in Fig.~\ref{q_opt_rs}. It should be emphasized that we do not find any optimal spin-spiral wave vector ${q_{opt}<k_f}$.
Note that for densities close to the transition to the 
ferromagnetic state the optimum wave vector $q_{opt}$ can be quite different 
from $2 k_f$ while for higher densities it is very close to this value. 

\begin{figure}
\includegraphics[clip,width=0.95\columnwidth]{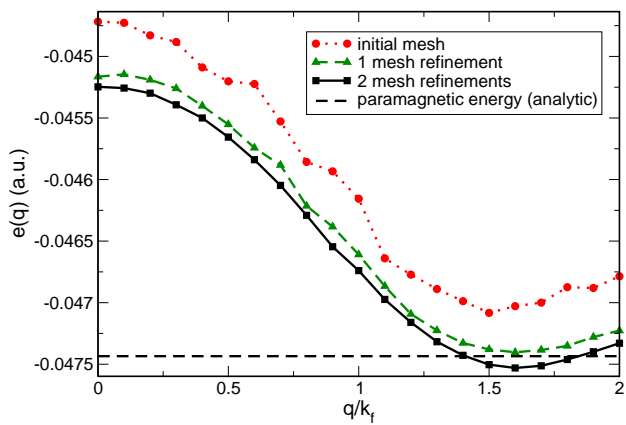}
\caption{\label{hf_e_rs5} 
(Color online) Hartree-Fock total energy per electron as function of the spin-spiral 
wave vector $ q $ at the density corresponding to ${r_s=5.0}$.
The data sets represent results at different discretizations. The dashed horizontal line visualizes the analytic
value of the paramagnetic ground state energy. The optimal spin-spiral wave vector is ${q_{opt}\approx1.6 k_f}$.} 
\end{figure}

The effect of the refinement of the discretization in momentum space is shown 
in Fig.~\ref{hf_e_rs5}. 
By sampling ${n_{b i}}$ and $a_i$ more often in regions of higher 
variations we both lower the energy and reduce the numerical noise in ${e\blr{q}}$.
The convergence of the total energy can 
be inferred from the values ${e\blr{q=2 k_f}}$ at different discretizations and
comparing to the analytic paramagnetic energy. For the case of ${r_s=5.0}$ we 
obtain a spin-spiral energy that is lower than the analytic paramagnetic 
energy at the optimal value of the spin-spiral wave vector.
At higher densities (lower $r_s$) the energy gain by forming a SSDW is lower, so we would need a 
very fine discretization to obtain numerical results lower than the analytic paramagnetic energy.
However, considering the numerical value of the paramagnetic energy at the same discretization is sufficient to
demonstrate the instability w.r.t.~a SSDW formation because the computed 
energies are variational as discussed in Sec.~\ref{sec_num}.
In order to determine the dependence of the optimal spin-spiral wave vector 
$q_{opt}$ on the density, we therefore refine
the momentum space discretization until $q_{opt}$ is converged.

For our numerical results we have verified that the ONs and the angular 
parameters $ a_i $ satisfy Overhauser's 
self-consistent equations \eqref{ss_hf_sc} and \eqref{theta_pot} by iterating them
only once. The difference between the angles $a_i$ 
in the occupied regions before and after the iteration is numerically zero for all values of $\vec{q}$.
This means that choosing a spin-spiral wave vector we 
can always find a solution of the self-consistent 
equations derived by Overhauser. 
\begin{figure}
\includegraphics[clip,width=0.95\columnwidth]{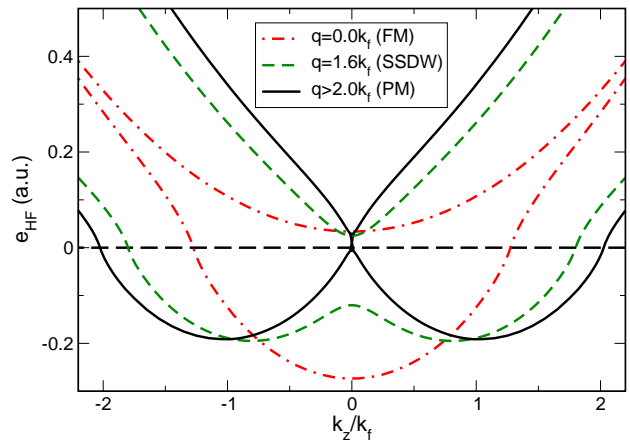}
\caption{\label{hf_sp_disp_rs5} 
(Color online) Hartree-Fock single-particle dispersion ${(k_{\rho}=0)}$ at ${r_s=5.0}$ for the paramagnetic
(${q\geq2k_f}$),
ferromagnetic (${q=0k_f}$) and
the SSDW state (${q_{opt}=1.6k_f}$). The single-particle energies are shifted such that the dashed horizontal line
corresponds to the Fermi energy for all $q$.
The difference between the two symmetric minima corresponds to the spin-spiral
wave vector $ q $. The paramagnetic dispersion may also be viewed as a 
spin-spiral dispersion with the origin in momentum space
shifted by ${\pm q}$ for the different spin channels [cf.~Eqs.~\eqref{pm_fm}].}
\end{figure}
Since the total energy does not depend on the $ a_i $ in regions where 
${n_{bi}=0}$, one self-consistency loop furthermore fixes the angles ${a_i}$ 
in unoccupied regions of $k$-space because they appear only on the left-hand 
side of Eqs.~\eqref{theta_pot}.
This is necessary to construct the proper HF dispersions (cf.~Fig.~\ref{hf_sp_disp_rs5}) also for the 
unoccupied states.
In a complementary work we have investigated the SSDW state using the
Optimized-Effective-Potential (OEP) method within the framework of non-collinear
Spin-Density-Functional Theory (SDFT)\cite{KurthEich:09}. In contrast to our findings within the OEP-DFT
framework, i.e.~, an effective single-particle theory restricted to local external potentials, here
we do not find holes below the Fermi surface (cf.~Ref.~\onlinecite{KurthEich:09} for details). This is
expected because it was shown in Ref.~\onlinecite{BachLieb:94} that the HF ground state has no holes below the
Fermi surface if the interaction is repulsive. Therefore our assumption of occupying only one band is justified.

\begin{figure}
\includegraphics[clip,width=0.95\columnwidth]{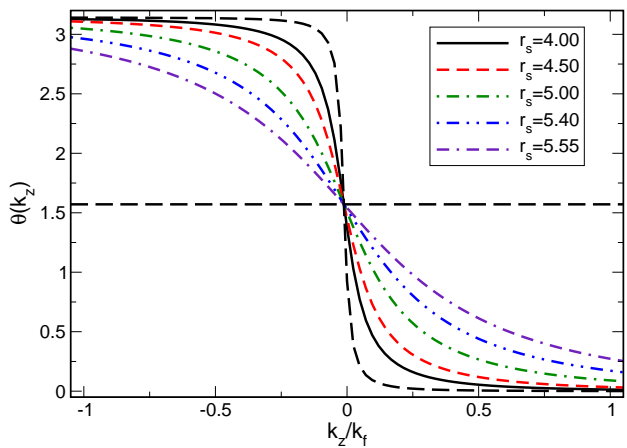}
\caption{\label{theta_var_rs} 
(Color online) Orbital angles ${\theta\blr{k_{\rho}=0,k_z}}$ for various densities,
specified by $r_s$, at the optimal spin-spiral wave vector.
The horizontal dashed line corresponds to the orbital angles at ${q=0}$ (ferromagnetic) and the step-like dashed line
corresponds to ${q=2k_f}$ (paramagnetic). For increasing $r_s$ the optimal spin-spiral wave vector becomes smaller,
such that the Fermi spheres, separated at ${q=2k_f}$, begin to overlap.
In order to gain energy the spin-up and spin-down orbitals
in the overlapping region hybridize and the orbital angle ${\theta}$ describes
the mixing of the spin-up and spin-down states.}
\end{figure}

At the single-particle level we have an intuitive understanding of the instability: as the two distinct spin-up
and spin-down regions of the paramagnetic state are squeezed into each other, 
the orbitals in the overlapping region hybridize. This hybridization then leads to the opening of a direct gap between the
HF single-particle dispersions corresponding to ${b=1,2}$ at $k_z=0$ as well 
as to a lowering of both the symmetry and the total energy of the system.
The mixing of the spin-up and spin-down orbitals is given by the orbital angles $\theta_{\vec{k}}$, capable of 
describing a continuous transition between the paramagnetic and the ferromagnetic state [Eqs.~\eqref{pm_conf} and \eqref{fm_conf}
respectively].
The behavior of the orbital angles at the optimal spin-spiral wave vector is shown in Fig.~\ref{theta_var_rs}.

\subsection{Correlated Functionals}
\begin{figure}
\includegraphics[clip,width=0.95\columnwidth]{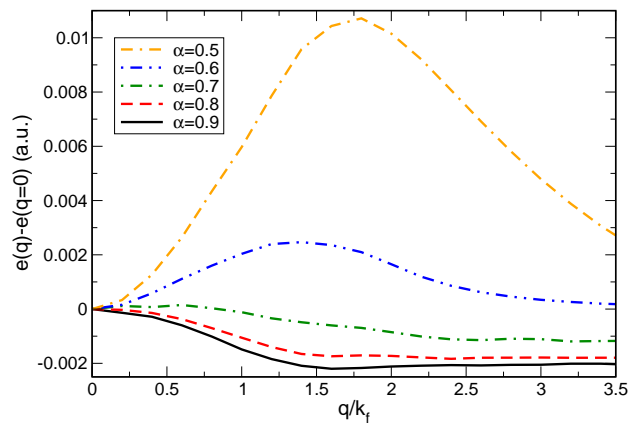}
\caption{\label{e_rs5_var_alpha} 
(Color online) Total energies per electron of the SSDW state described with the 
density-matrix-power functional as a function of the spin-spiral 
wave vector $q$ for various values of $\alpha$ at ${r_s=5.0}$.
The total energy per electron
at ${q=0}$ is subtracted in order to emphasize the behavior with increasing $q$.}
\end{figure}

The density-matrix-power functional reduces to the uncorrelated HF approximation 
for ${\alpha=1}$ and to the M\"uller functional
for ${\alpha=0.5}$. The latter one is known \cite{LathiotakisGross:07} to 
over-correlate and therefore 
one expects that decreasing $\alpha$ from ${1\to0.5}$ increases the amount of 
correlation in the system. This picture was verified in
Ref.~\onlinecite{LathiotakisGross:09}, where an optimal value of ${\alpha\approx0.6}$ was found in the
regions of metallic densities for the paramagnetic UEG.
In Fig.~\ref{e_rs5_var_alpha} the dependence of the total energy per 
particle at ${r_s=5.0}$ is shown for various $\alpha$.
It should be noted that the configuration for ${q>2k_f}$ cannot be interpreted 
as the paramagnetic state in the correlated case. This is due to the fact that 
correlations smear out the sharp step found for the uncorrelated case in the 
momentum distribution around the Fermi surface (see
 Ref.~\onlinecite{LathiotakisGross:07} for details). Therefore at ${q=2k_f}$ the (fractionally)
occupied regions in momentum space are not necessarily disjoint.
Only when the occupied regions separate into two parts the configuration may corresponds to the
paramagnetic state. However, the configuration at ${q=0}$ may still be 
interpreted as the ferromagnetic state.
\begin{figure}
\includegraphics[clip,width=0.95\columnwidth]{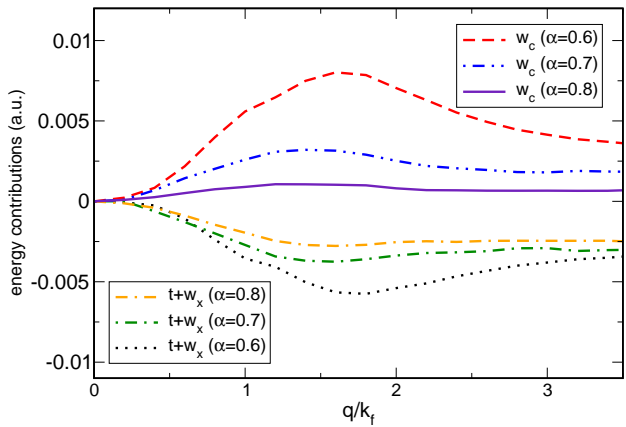}
\caption{\label{e_cont_rs5} 
(Color online) Comparison of the correlation energy with the 
contribution from kinetic plus exchange terms (KEX)
for ${\alpha=0.6,0.7,0.8}$ at ${r_s=5.0}$. All energy contributions are shifted such that the value at ${q=0}$ is zero. For
decreasing $\alpha$ the minimum in the KEX contribution shifts to higher values of $q$. The correlation contribution
however damps out this instability for values of $\alpha$ that yield good total energies at metallic densities.}
\end{figure}

From Fig.~\ref{e_rs5_var_alpha} it is clear that the instability w.r.t.~a SSDW is still present for 
${\alpha=0.9}$. For higher values of $\alpha$ the instability disappears and for ${\alpha=0.5,0.6}$ the energy has a maximum
in the SSDW region. Thus for values of $\alpha$ which provide good correlation energies for the UEG in the paramagnetic regime
there is no SSDW formation. In order to understand the reason for this it is instructive to look at various contributions to 
the total energy. In Fig.~\ref{e_cont_rs5} we compare the correlation energy contribution with the contribution coming
from the kinetic and exchange terms. The minimum is still present considering only kinetic and exchange contributions, but
for decreasing ${\alpha}$ the correlation contribution damps out the instability more and more.
One might suspect that at high densities, where exchange dominates correlations, the
instability sustains. Our findings in Sec.~\ref{sec_hf} show that in the HF approximation the 
energy gain decreases when the density increases, which is consistent with an analytic 
argument \cite{GiulianiVignale:08} that at high densities the energy gain 
by forming a SDW and/or CDW is overcome by correlations. 
Furthermore our results indicate that correlation effects
dominate the SSDW instability also at intermediate densities.

\begin{figure}
\includegraphics[clip,width=0.95\columnwidth]{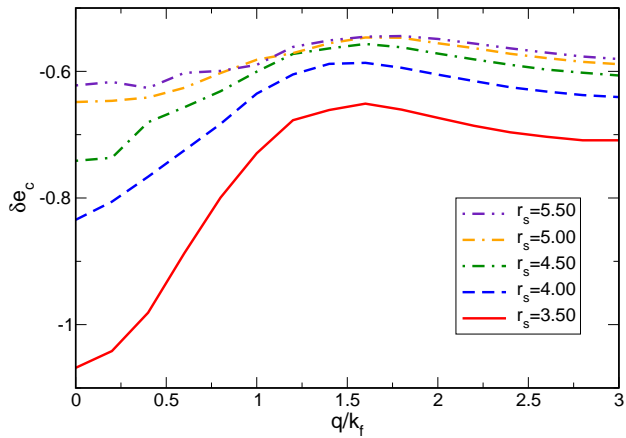}
\caption{\label{dec_bba0.6}
(Color online) Relative correlation energy as a function of the spin-spiral wave vector $q$ for various $r_s$ at ${\alpha=0.6}$.
Correlations are more important around the ferromagnetic configuration. In the region of the Hartree-Fock SSDW 
instability correlations have the smallest effect. This indicates why the instability is not sustained
when correlations are included at the level of the density-matrix-power functional. 
}
\end{figure}
In order to gain some insight into the role
of correlations we define the relative correlation energy $\delta e_c$ as 
\begin{equation} \label{dec_def_1}
\delta e^{\alpha}_c \equiv \frac{w_{\alpha}-w_{HF}}{\abs{e_{\alpha}}} \; .
\end{equation}
In Fig.~\ref{dec_bba0.6} we show the dependence of this quantity on the 
spin-spiral wave vector for the correlation 
parameter ${\alpha=0.6}$. The absolute value of the relative correlation is 
smallest in the region of 
the SSDW instability (${q=k_f\to2k_f}$), which explains why the instability is no longer present 
when correlations are included. Furthermore we can see that the relative 
correlation is dominant in the region of the ferromagnetic configuration. 
This can be understood by noticing that the density-matrix-power functional approximates the 
correlation energy by a prefactor times a Fock integral (most present-day functionals in RDMFT approximate correlations 
in this way\cite{Kollmar:04,CioslowskiBuchowiecki:03,CioslowskiPernal:05,CsanyiArias:00,CsanyiArias:02,
BuijseBaerends:02,GritsenkoBaerends:05,GoedeckerUmrigar:98,Piris:05,LathiotakisGross:07,MarquesLathiotakis:08}).
Since Fock integrals imply that equal spins
are particularly correlated, one would expect a similar dependence of the relative correlation energy
for other RDMFT functionals.

\section{Summary and Conclusion}
We have investigated the instability of the uniform electron gas w.r.t.~the 
formation of a spin-spiral density wave within Reduced-Density-Matrix-Functional Theory, which includes the 
Hartree-Fock approximation as an important limiting case. To our knowledge 
this is the first numerical Hartree-Fock study of the spin-spiral state in the electron 
gas, despite the fact that Overhauser presented his analytical work on the 
problem more than four decades ago. In Overhauser's work, the optimal
spin-spiral wave vector was not determined. Our study shows that, in contrast to common belief, the optimal 
spin-spiral wave vector is not always close to ${2 k_f}$. While at high densities we confirm
this value for the optimal wave vector, for lower densities (just before the transition to the ferromagnetic state)
the optimal wave vectors even approaches ${k_f}$.

Within the framework of Reduced-Density-Matrix-Functional Theory we also 
studied the effect of correlations on the spin-spiral density wave instability 
using the recently proposed density-matrix-power functional.
Not unexpectedly, we find that the inclusion of correlations suppresses the 
instability, which is explained by the behavior of the correlation energy 
in the region of the spin-spiral density wave instability.

\acknowledgments

We would like to acknowledge useful discussions with Giovanni
Vignale. We also acknowledge funding by the "Grupos Consolidados
UPV/EHU del Gobierno Vasco" (IT-319-07). C.~R.~P.~was
supported by the European Community through a Marie Curie
IIF (Grant No.~MIF1-CT-2006-040222) and CONICET of
Argentina through Grant No.~PIP 5254.

\end{document}